\documentclass{stsci_report}

\usepackage{graphicx}
\usepackage{listings}
\usepackage{xcolor}

\usepackage{siunitx}
\usepackage{todonotes}
\usepackage{pdfpages}
\usepackage{hyperref}
\usepackage{parskip}
\usepackage{dirtytalk}
\usepackage[section]{placeins}
\usepackage{caption}
\usepackage{aas_macros}

\usepackage{float}
\usepackage{graphicx}
\usepackage{siunitx}
\usepackage{todonotes}
\usepackage{pdfpages}
\usepackage{tabularx}
\usepackage{multirow}
\usepackage{longtable}
\usepackage{caption}
\usepackage{subcaption}
\usepackage{aas_macros}
\usepackage[bottom]{footmisc}
\usepackage{wrapfig}
\usepackage{amsmath}
\usepackage{colortbl}
\usepackage{tabularray }
\usepackage{setspace} 
\usepackage{natbib}
\usepackage{listings}
\usepackage{ragged2e}

\newcommand{\specialcell}[2][c]{%
  \begin{tabular}[#1]{@{}c@{}}#2\end{tabular}}

\DeclareCaptionType{equ}[][]
\usepackage{amsmath} 

\definecolor{codegreen}{rgb}{0,0.6,0}
\definecolor{codegray}{rgb}{0.5,0.5,0.5}
\definecolor{codepurple}{rgb}{0.58,0,0.82}
\definecolor{backcolour}{rgb}{0.95,0.95,0.92}

\lstdefinestyle{mystyle}{
  backgroundcolor=\color{backcolour},   
  commentstyle=\color{blue},
  keywordstyle=\color{codegreen},
  numberstyle=\tiny\color{codegray},
  stringstyle=\color{codepurple},
  basicstyle=\ttfamily\footnotesize,
  breakatwhitespace=false,         
  breaklines=true,                 
  captionpos=b,                    
  keepspaces=true,                 
  numbers=left,                    
  numbersep=5pt,                  
  showspaces=false,                
  showstringspaces=false,
  showtabs=false,                  
  tabsize=1
}

\copyrighttext{Copyright\copyright\ \the\year\ The Association of Universities for Research in Astronomy, Inc. All Rights Reserved.}

\presubtitle{Instrument Science Report WFC3 2025-02}
\title{WFC3/UVIS EPER CTE 2009-2025}
\author{Anne O'Connor, Harish Khandrika}

\date{May 28, 2025}

\begin{document}

\maketitle

\abstract{In this report, we examine the behavior of Charge Transfer Efficiency (CTE) on the WFC3/UVIS detector over time as computed by the Extended Pixel Edge Response (EPER) technique, using internal calibration data acquired from 2009 through 2025. We find that the CTE has continued to decline as expected, with a steeper loss rate for lower signal levels. The lowest signal level (160~e\text{-}) has continued to decline at a rate of 0.0001 per year, with a total overall decline of 0.0015. Analyses from 2016 and 2020 found that the rate of decline was not well fit by a linear function. This report verifies the rate of decline is instead better fit by a quadratic function (which results in the smallest min. and max. residuals, on average) or a cubic function (which has the best \say{goodness of fit} $\chi^2$ and $R^2$ values). We continue to see periodic oscillations of the residuals around all three fit lines (linear, quadratic, and cubic) on which we perform a Lomb-Scargle periodogram analysis of the residuals. We find a periodicity of about 8 years for the residuals around the linear fit lines and about 9 years for the quadratic and cubic fit lines.}

\section{Introduction}
All HST CCD detectors, including WFC3/UVIS, experience degradation to the Charge Transfer Efficiency (CTE) of the detector, mainly caused by continuing damage to the silicon lattice by cosmic rays \citep{2016wfc..rept...10K}. CTE losses can degrade the precision of stellar photometry and astrometry measurements taken with the affected CCD detectors, so it is important to monitor and characterize the degradation. One method of monitoring CTE loss in the WFC3/UVIS channel is the Extended Pixel Edge Response (EPER) technique, initially applied to WFC3 data during the April 2007 Ambient Calibration campaign \citep{2007wfc..rept...13R}. The EPER technique, described by \cite{2001sccd.book.....J}, takes advantage of a special readout mode that provides a larger-than-normal overscan area (extended pixel region). The technique measures the excess charge in the overscan pixels of the CCD, which appears as an exponential tail of charge that decreases with distance from the science pixels (described in more detail in \cite{2009wfc..rept...10K}) . The most recent EPER-based analysis of CTE losses in the WFC3/UVIS detector, conducted in 2020, found that the CTE had declined below  0.9990 for the lowest signal level (160~e\text{-}) over the previous 10 years at a rate of 0.0001 per year \citep{2020wfc..rept....6K}. Additionally, \cite{2020wfc..rept....6K} found that the residuals of the linear fit exhibit a periodicity that is anti-correlated with solar activity.

This report serves as an update to the previous WFC3/UVIS EPER CTE measurements (\cite{2020wfc..rept....6K} and references therein), and presents an analysis and results of WFC3/UVIS EPER observations acquired for all data from launch in 2009 (HST Cycle 17) to 2024-2025 (HST Cycle 32).

\section{Data}

 The structure of the visits and the exposure parameters for data taken in the WFC3/UVIS EPER calibration programs has remained the same across all programs and cycles, which provides a stable dataset from which to measure the EPER CTE. For each program, short internal flat-field observations are taken in pairs with two different filters at various exposure times to achieve specific illumination levels. Table \ref{tab:obs_param} outlines the filter, exposure time, illumination level, and image type for each exposure in the pair of EPER visits. Table \ref{tab:prop_list} lists the calibration programs through which the UVIS EPER data were acquired. 

\begin{minipage}[!b]{0.9\linewidth} \vspace{3ex}
\normalsize
    \centering 
    \def\arraystretch{1.3} 
    \captionsetup{type=table}
    \captionof{table}{ Observational parameters for a two-visit pair, where n is the first visit in the series and n+1 is the second.}
    \begin{tabular}{|c|c|c|c|c|} \hline
\textbf{Visit}  & \textbf{Image Type} & \textbf{Filter} & \textbf{Exp. Time} & \textbf{Illumination Level}  \\ \hline
         n    & DARK & -          & 0.5 &  - \\ \hline 
         n    & TUNGSTEN & F390M  & 9.2 &  $160e\text{-}$ \\ \hline 
         n    & TUNGSTEN & F390M  & 22.9&  $400e\text{-}$ \\ \hline 
         n + 1& DARK &       -    & 0.5 &  - \\ \hline 
         n + 1& TUNGSTEN &  F390W & 6.4 &  $800e\text{-}$ \\ \hline
         n + 1& TUNGSTEN &  F438W & 7.6 &  $1600e\text{-}$ \\ \hline
         n + 1& TUNGSTEN &  F438W & 22.7&  $5000e\text{-}$ \\ \hline

    \end{tabular}
    
    \label{tab:obs_param}
\vspace{2ex}
\end{minipage}

\begin{minipage}[!b]{0.9\linewidth} \vspace{3ex}
\normalsize
    \centering 
    \def\arraystretch{1.3} 
    \captionsetup{type=table}
    \captionof{table}{ List of calibration proposals to measure CTE loss in WFC3/UVIS using the EPER method.}
    \begin{tabular}{|c|c|c|c|} \hline
\textbf{Proposal ID}  & \textbf{Cycle} & \textbf{Principal Investigator} & \textbf{Frequency of Obs.}  \\ \hline
         11924& 17 & Kozhurina-Platais &  Once a month \\ \hline 
         12347& 18 & Kozhurina-Platais &  Once a month \\ \hline 
         12691& 19 & Kozhurina-Platais &  Once a month \\ \hline 
         13082& 20 &  Bourque          &  Once a month \\ \hline 
         13565& 21 &  Bourque          &  Every other month \\ \hline
         14011& 22 &  Bowers           &  Every other month \\ \hline
         14377& 23 &  Khandrika        &  Every other month \\ \hline
         14540& 24 &  Khandrika        &  Every other month \\ \hline
         14989& 25 &  Fowler           &  Every other month \\ \hline
         15575& 26 &  Fowler           &  Every other month \\ \hline
         15720& 27 &  Fowler           &  Every other month \\ \hline
         16400& 28 & Khandrika         & Every other month \\ \hline
         16572& 28 & Khandrika         & Every other month \\ \hline
         17008& 30 & Khandrika         & Every other month \\ \hline
         17353& 31 & Khandrika         & Every other month \\ \hline
         17673& 32 & O'Connor          & Every other month \\ \hline

    \end{tabular}
    
    \label{tab:prop_list}
\vspace{2ex}
\end{minipage}

\section{Analysis}
An IDL script, as described in \cite{2011wfc..rept...17K} with updates and further details in \cite{2016wfc..rept...10K}, was developed to measure the CTE loss from the on-orbit EPER observations and used in subsequent analyses. The full methodology of the EPER analysis is described in detail in the aforementioned reports, but a brief summary of the analysis is as follows: internal flat fields are taken in a special readout mode which generates extra overscan areas, and the deferred charge due to CTE within that overscan (or extended pixel region) is measured. The overscan columns are first averaged to mitigate random noise and increase the signal-to-noise ratio, and any extraneous signal (i.e. due to low-level periodic electronic noise that varies from image to image) is removed via sigma-clipping. The CTE is then calculated as the ratio of the total deferred charge in the overscan column average to the charge level in the last column of science pixels multiplied by the number of pixel transfers in the CCD register. We followed a similar procedure in this report, using the legacy IDL script (mentioned above) to process our data. 

\section{Results}

\begin{figure}[!t]
    \centering
    \includegraphics[width=18cm]{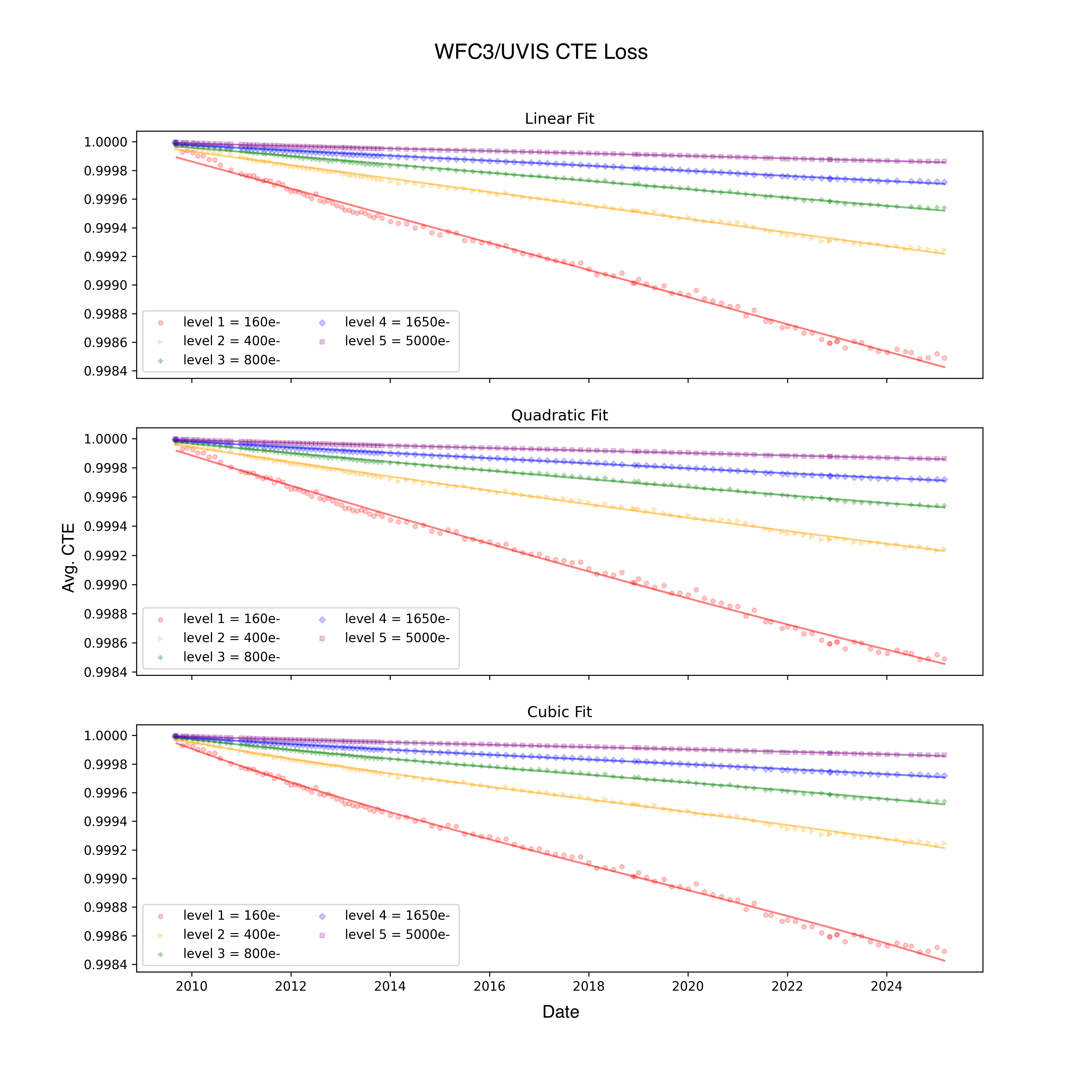}\caption{Average CTE measurement using the EPER method as a function of time for 5 different illumination levels: 160e-, 400e-, 800e-, 1650e-, and 5000e-. A linear (top), quadratic (middle), and cubic (bottom) function are fitted to the data. The WFC3/UVIS CTE as measured by the EPER method continues to decline with time and declines more steeply at fainter illumination levels.
}
    \label{fig:CTE_losses}
\end{figure}

Figure \ref{fig:CTE_losses} shows the CTE as a function of time and illumination level. We find that the WFC3/UVIS CTE, as measured using the EPER method, has continued to decline with time and declines more steeply at fainter illumination levels, as observed in previous studies \citep{2020wfc..rept....6K}. The rate of decline in CTE per year and the overall level of decline in the CTE per illumination level is listed in Table \ref{tab:CTE_decline}. For the lowest signal level, the CTE has declined by up to 0.0015  over the last 15 years with a rate of decline of 0.0001 per year.

Figure \ref{fig:CTE_losses} shows that  CTE values are closer to the ideal value of 1.0 at higher signal levels, which suggests a power law relationship between CTE and signal level. \cite{2009wfc..rept...10K} discusses this, and defines the power-law relationship between the CTE and signal level with Equation \ref{eqn:power_law_relationship}: 

 \begin{equ}[h]
\captionsetup{width=.9\linewidth}
\begin{equation}
\centering
\label{eqn:power_law_relationship}
\begin{aligned}
CTE = 1 - m \times S^\rho
\end{aligned}
\end{equation}
\caption*{Equation 1: Power law relationship between CTE and signal level, where $m$ and $\rho$ are  free parameters and $S$ is the signal level of the last column (in electrons).} 
\end{equ} 

Another way to quantify CTE loss is charge transfer inefficiency (CTI), which can be defined as $CTI = 1 - CTE$. We can then rewrite Equation \ref{eqn:power_law_relationship} as:
 \begin{equ}[h]
\captionsetup{width=.9\linewidth}
\begin{equation}
\centering
\label{eqn:log_relationship}
\begin{aligned}
log(CTI) = log(m) + \rho\times log(S)
\end{aligned}
\end{equation}
\caption*{Equation 2: The relationship between CTI (the loss of charge as signal is transferred between pixels) and signal level (S), where $log(m)$ and $\rho$ are the slope and the intercept of the log-log CTI, respectively.} 
\end{equ} 

\begin{figure}[!t]
    \centering
    \includegraphics[width=18cm]{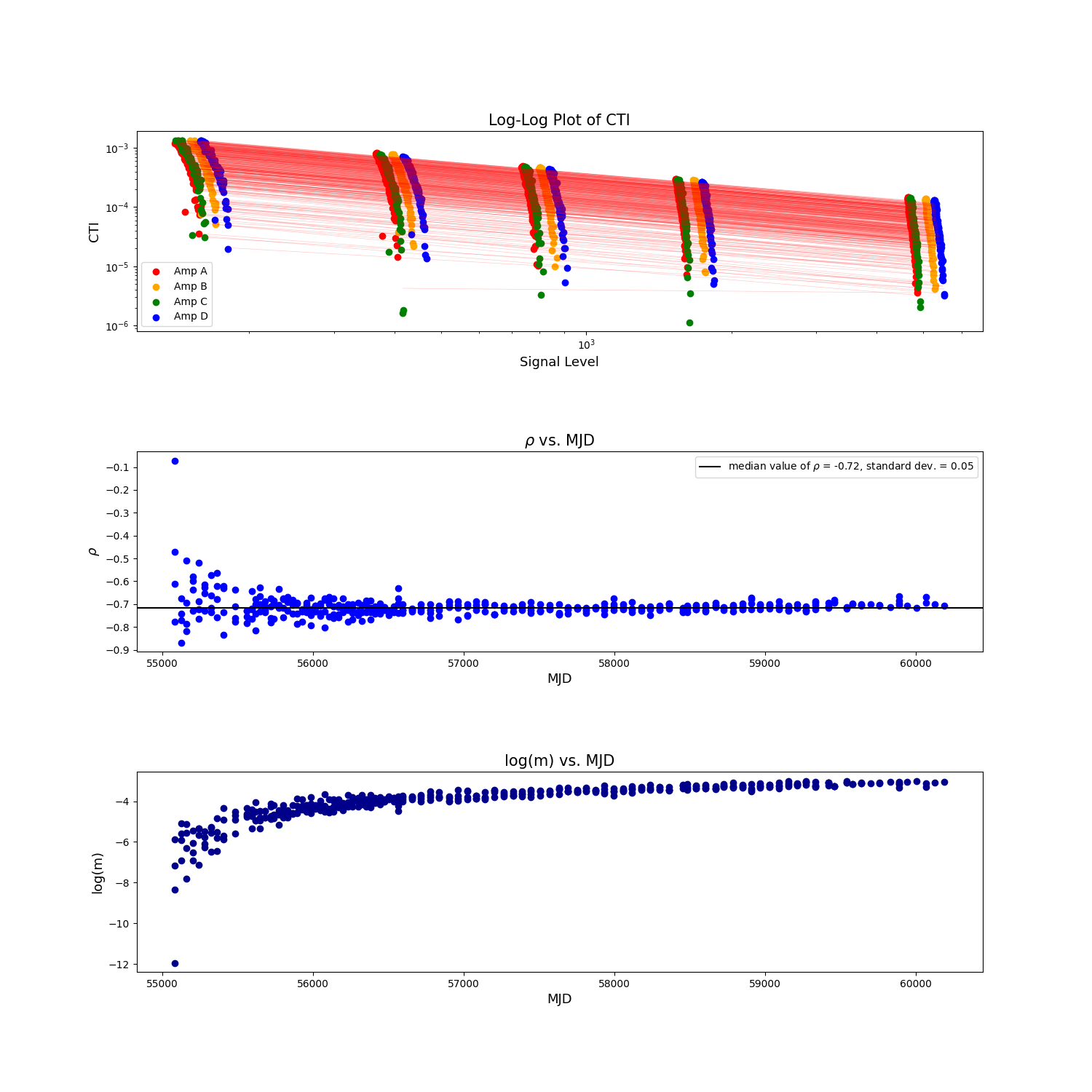}\caption{(Top) CTI versus the signal level in electrons with a linear fit (overplotted in red) to CTI values for images within each observation. Four different colors represent each of the four amps of the detector. (Middle) $\rho$ (from the power law relationship of CTI to signal level) versus time on a log-log scale. (Bottom) The intercept of the CTI ($log(m)$) as a function of time on a log-log scale.}
    \label{fig:Log-CTI}
\end{figure}

Figure \ref{fig:Log-CTI} shows the measured EPER CTI as a function of signal level, with linear fits for each observation and colors to represent the CTI value for each of the detector's amps (A, B, C, and D). The slope ($\rho$) and intercept $log(m)$ for the CTI, described in Equation \ref{eqn:log_relationship}, are plotted in Figure \ref{fig:Log-CTI} as well.  We find a stable CTI slope over time where $\rho = -0.72 \pm 0.05$ (where the uncertainty is reported as the standard deviation of the $\rho $ values) and an increasing intercept ($log(m)$) over time. These results are consistent with \cite{2016wfc..rept...10K} and the previous reports mentioned therein, which report $\rho = -0.68$ and an increasing CTE intercept over time.

\begin{minipage}[!b]{0.9\linewidth} \vspace{3ex}
\normalsize
    \centering 
    \def\arraystretch{1.3} 
        
    \captionsetup{type=table}
    \captionof{table}{Overall decline in the CTE with time for all illumination levels, corresponding
    to the points seen in Figure \ref{fig:CTE_losses}. A linear fit to each of the illumination levels was made (as seen in the top plot of Figure \ref{fig:CTE_losses}) and
    the slope and uncertainty in slope are shown here as the linear decline in CTE level per year.}
 \begin{tabular}{|c|c|c|c|}
\hline
\specialcell{\textbf{Illumination} \\ \textbf{Level (e\text{-})}} & \specialcell{ \textbf{Overall CTE} \\ \textbf{Decline}} & \specialcell{\textbf{Linear Decline
} \\ \textbf{per Year (Slope)}} &\specialcell{\textbf{Uncertainty} \\ \textbf{in Slope}} \\
\hline
160~e$\text{-}$   & 0.00151 & $-9.57 \times 10^{-5}$& $\pm 0.08 \times 10^{-5}$  \\
\hline
400~e$\text{-}$   & 0.00075 & $-4.76 \times 10^{-5}$ & $\pm 0.04 \times 10^{-5}$ \\
\hline
800~e$\text{-}$   & 0.00046 & $-2.91 \times 10^{-5}$ & $\pm 0.02\times 10^{-5}$ \\
\hline
1650~e$\text{-}$  & 0.00028 & $-1.78 \times 10^{-5}$ & $\pm 0.01 \times 10^{-5}$ \\
\hline
5000~e$\text{-}$  & 0.00014 & $-0.87 \times 10^{-5}$ & $\pm 0.01\times 10^{-5}$
\\
\hline
\end{tabular}

    \label{tab:CTE_decline}
\vspace{2ex}
\end{minipage}

\clearpage
\begin{figure}[!t]
    \centering
    \includegraphics[width=18cm]{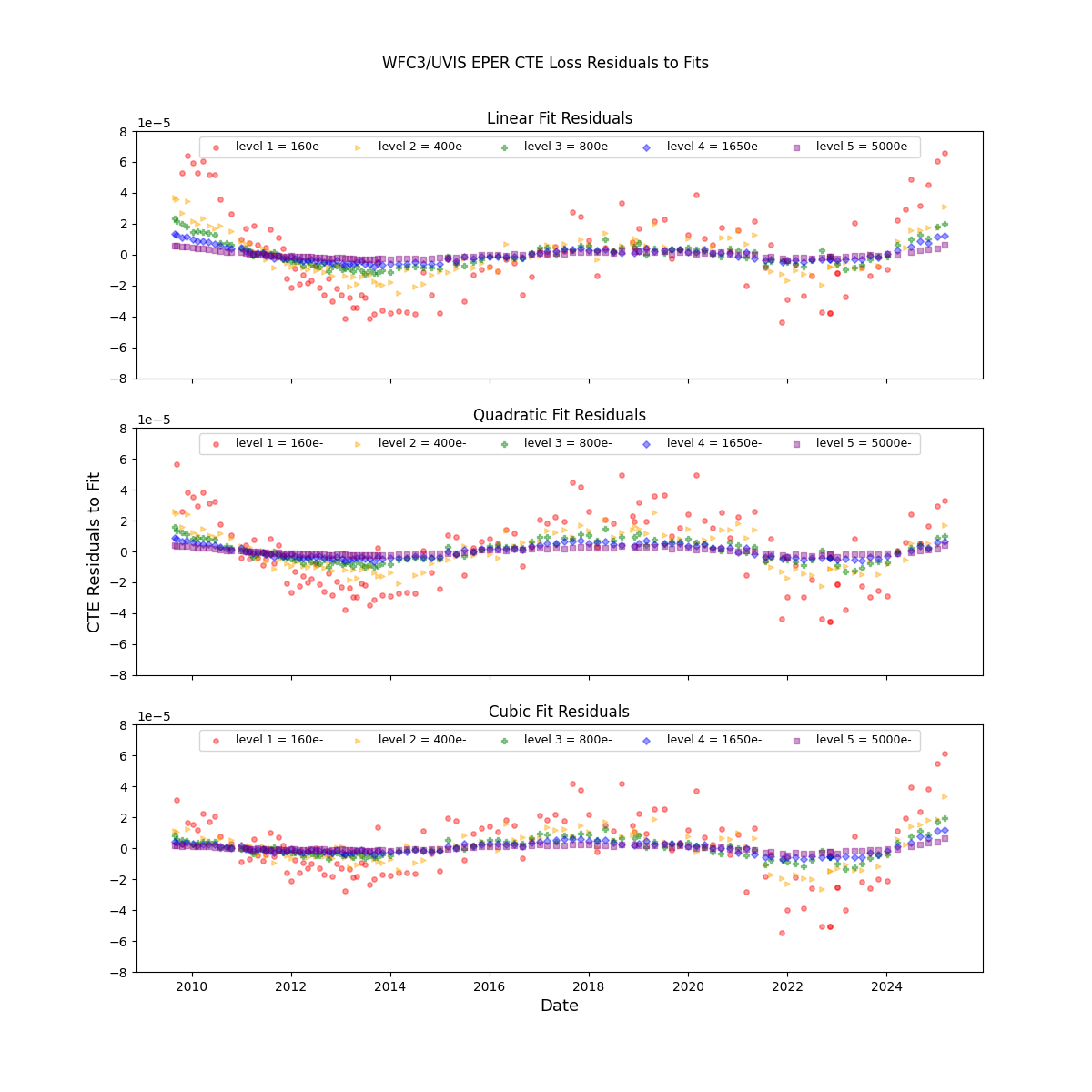}\caption{Residuals between the data and the linear (top), quadratic (middle), and cubic (bottom) fit functions for all illumination levels.}
    \label{fig:CTE_resids}
\end{figure}

\clearpage

\clearpage
To assess whether CTE was declining linearly, \cite{2020wfc..rept....6K} performed a linear, quadratic, and cubic fit to the CTE data, and found that the quadratic fit resulted in the lowest reduced chi-squared and smallest residuals. We re-confirm the 2016 and 2020 findings that the rate of decline is not well-matched by a linear fit. We provide the reduced $\chi^2$ goodness-of-fit values, coefficient of determination values ($R^2$), maximum residuals, and minimum residuals in Table \ref{tab:chi-R-sq}. We find that the data are better fit by a quadratic function (which has the smallest min. and max. residuals, on average) or cubic function (with the smallest reduced $\chi^2$ and best $R^2$ values) .

\begin{minipage}[!b]{0.9\linewidth} \vspace{3ex}
\normalsize
    \centering 
    \def\arraystretch{1.3} 
    \captionsetup{type=table}
    \captionof{table}{Various fits, residuals, reduced $\chi^2$ and the coefficient of determination values ($R^2$) for the different illumination levels.}
    \begin{tabular}{|c|c|c|c|c|c|} \hline
\specialcell{\textbf{Illumination} \\ \textbf{Level}}& \textbf{Fit} & \textbf{Reduced $\chi^2$} & \textbf{$R^2$} &
\specialcell{\textbf{Residual Min} \\ (x$10^{-5}$)} & \specialcell{\textbf{Residual Max} \\ (x$10^{-5}$)}\\ \hline
      160~e\text{-}   & Linear & 3.91 & 0.9960 & -4.3&  8.5\\ \hline
      160~e\text{-}   & Quadratic & 2.85 & 0.9969 & -4.5&  5.7\\ \hline
      160~e\text{-}   & Cubic & 2.21 &  0.9975 &-5.5&  6.1\\ \hline \hline
      
      400~e\text{-}   & Linear & 5.84 & 0.9965 & -2.5&  3.7\\ \hline 
      400~e\text{-}   & Quadratic  & 4.47 & 0.9972 & -2.2&  2.6\\ \hline 
      400~e\text{-}   & Cubic & 3.11 & 0.9980 &-2.6&  3.4\\ \hline \hline
      800~e\text{-}   & Linear & 9.34  & 0.9965 & -1.2&  2.4\\ \hline 
      800~e\text{-}   & Quadratic & 6.87  & 0.9974 & -1.3&  1.6\\ \hline 
      800~e\text{-}   & Cubic & 4.80 & 0.9980 & -1.3&  2.0\\ \hline \hline
      1650~e\text{-}  & Linear & 13.50 & 0.9965 & -0.8&  1.4\\ \hline 
      1650~e\text{-}  & Quadratic & 9.81 & 0.9974 & -0.6&  0.9\\ \hline
      1650~e\text{-}  & Cubic &  6.90  & 0.9980 & -0.7&  1.2\\ \hline \hline
      5000~e\text{-}  & Linear & 25.69 & 0.9969 & -0.3&  0.6\\ \hline
      5000~e\text{-}  & Quadratic & 20.10 & 0.9975 & -0.4&  0.4\\ \hline
      5000~e\text{-}  & Cubic & 14.35 & 0.9981 & -0.5&  0.7\\ \hline

    \end{tabular}

    \label{tab:chi-R-sq}
\vspace{2ex}
\end{minipage}

\\

\begin{figure}[!t]
    \centering
    \includegraphics[width=18cm]{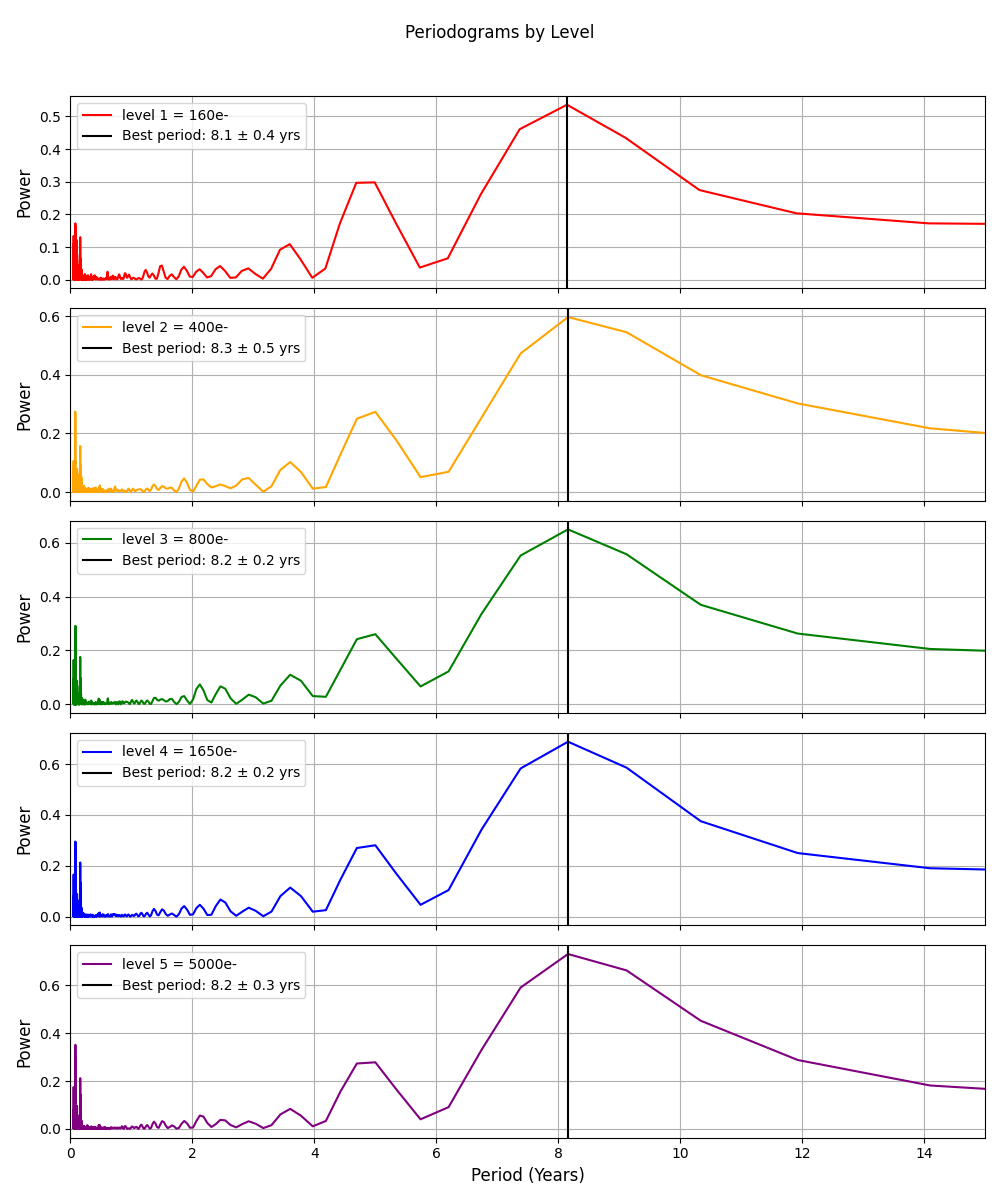}\caption{Lomb-Scargle periodogram of CTE-loss residuals around a linear fit line for each of the illumination levels. The characteristic best period is marked with a vertical black line.}
    \label{fig:lombscargle_linear}
\end{figure}

\cite{2020wfc..rept....6K} observed a cyclical nature of the linear fit residuals and found that it was inversely correlated with solar activity. We build upon the analysis and investigate cyclical oscillations in the linear, quadratic, and cubic fit residuals with additional data from 2009 to 2025, as shown in Figure \ref{fig:CTE_resids}.  We choose to analyze the residual oscillations around each of the fit lines (linear, quadratic, and cubic), and use \texttt{astropy} to compute the period of these oscillations by creating a Lomb-Scargle periodogram.\footnote{See \href{https://docs.astropy.org/en/stable/api/astropy.timeseries.LombScargle.html\#astropy.timeseries.LombScargle}{\texttt{astropy.timeseries.lombscargle}} documentation and Appendix A for more info.} Figure \ref{fig:lombscargle_linear} shows the periodograms of the residuals around the linear fit line for all five illumination levels. Figures \ref{fig:lombscargle_quad} and \ref{fig:lombscargle_cubic}, located in Appendix A, show the periodograms for all 5 illumination levels for residuals around the quadratic and cubic fit lines, respectively. 

We find that the CTE loss residuals oscillate at a period of about $8.1 \pm 0.4$ years around the linear fit line for an illumination level of 160~e\text{-} (please see Figure \ref{fig:lombscargle_linear} for all illumination levels) and at periods of about $9.1 \pm 0.3$ and $9.0 \pm 0.3$ around the quadratic and cubic fit lines (for an illumination level of 160~e\text{-}; see Appendix A for plots), respectively. Though the periodicity in the residuals is slightly smaller than the 11 year solar activity period, we are limited by the 15 year lifetime of the WFC3 instrument and we are likely underestimating the uncertainty in the periodicity. Thus, we cannot say conclusively whether or not the CTE loss residuals are correlated with solar activity.

\clearpage

\section{Conclusions}

We find that the WFC3/UVIS CTE loss as measured using the EPER method continues to decrease, with a steeper rate of decline at lower signal levels. The lowest illumination level of 160~e\text{-} declines the most at a rate of 0.0001 per year, which agrees well with previous results \citep{2020wfc..rept....6K}. We 
 also find that the rate of decline is better fit by a quadratic or cubic function compared to a linear function, thus confirming results from 2016 and 2020 (\cite{2016wfc..rept...10K}, \cite{2020wfc..rept....6K}). Finally, we continue to observe a cyclical pattern in the residuals, with a periodicity of about 8 or 9 years, depending on the fit function. CTE loss will continue to be monitored using the EPER method to identify any changes in behavior that may affect the health of the instrument.

\section{Acknowledgments}
 The authors would like to thank Ky Huynh, Isabel Rivera, Sylvia Baggett, and Joel Green for their detailed reviews of this report, whose feedback and suggestions helped improve this report. 

\bibliography{ref}
\bibliographystyle{aasjournal}
\nocite{*}

\section{Appendix A}

\begin{figure}[!t]
    \centering
    \includegraphics[width=18cm]{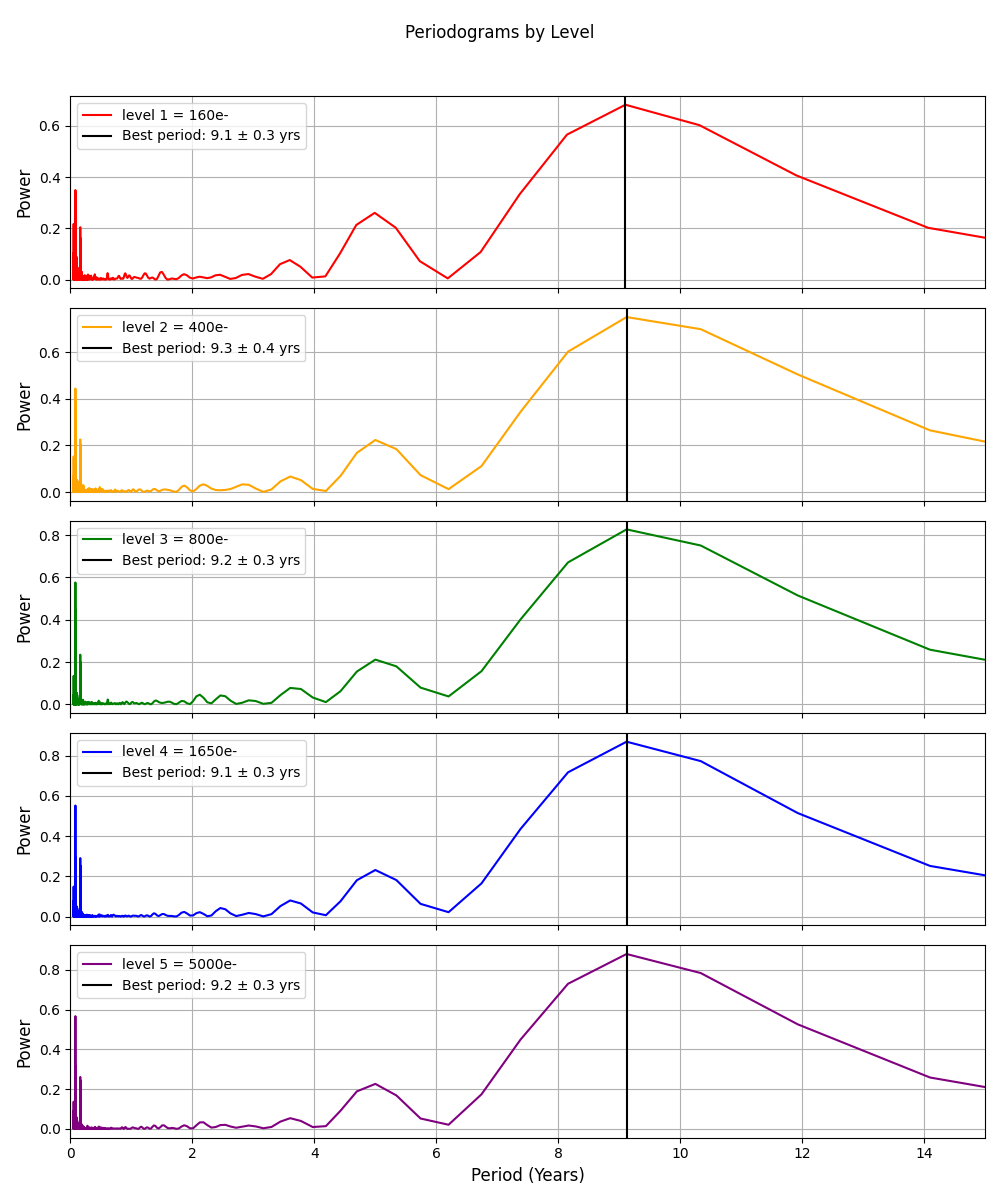}\caption{Lomb-Scargle periodogram of CTE-loss residuals around a quadratic fit line for each of the illumination levels. The characteristic best period is marked with a vertical black line.}
    \label{fig:lombscargle_quad}
\end{figure}

\begin{figure}[!t]
    \centering
    \includegraphics[width=18cm]{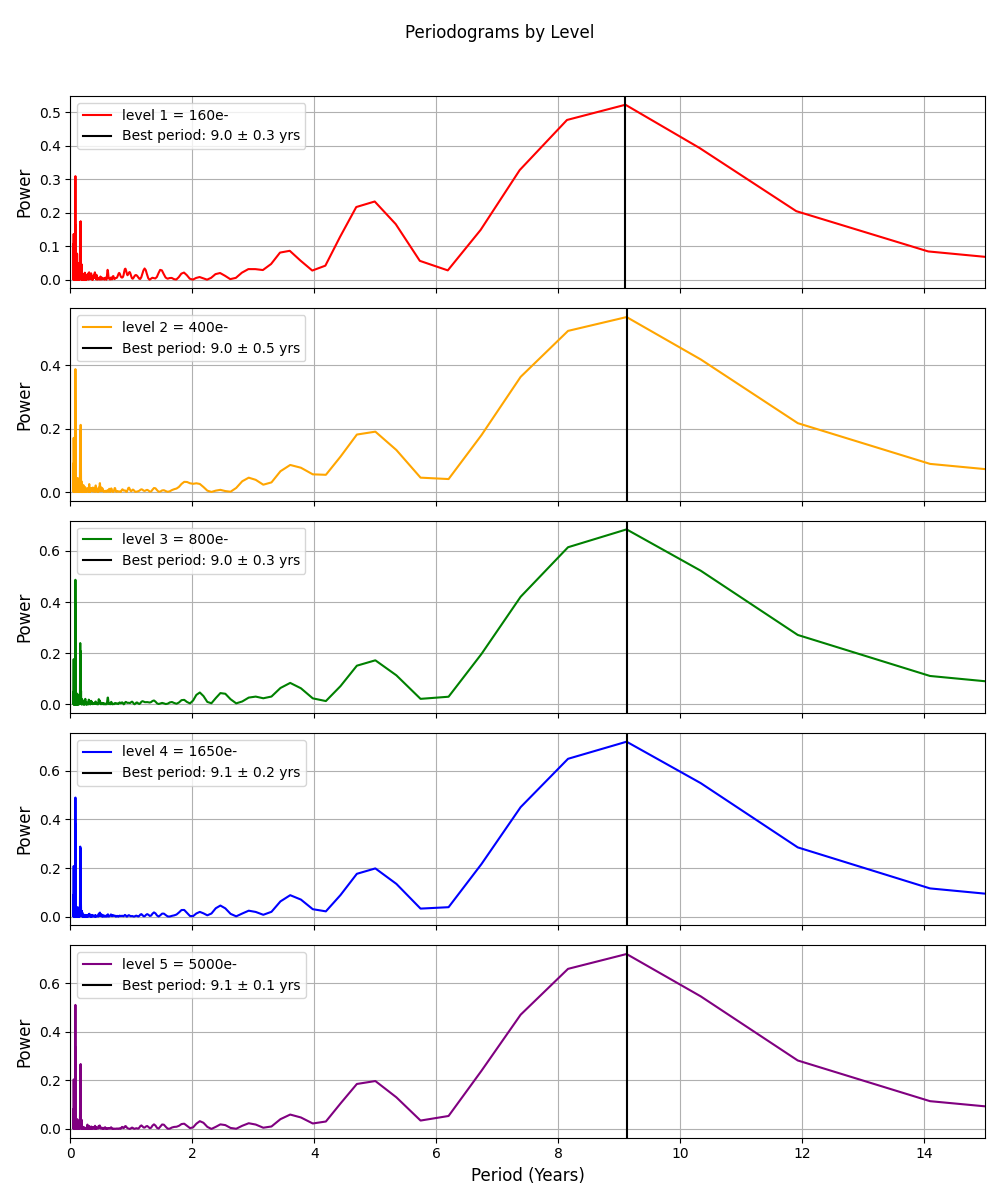}\caption{Lomb-Scargle periodogram of CTE-loss residuals around a cubic fit line for each of the illumination levels. The characteristic best period is marked with a vertical black line.}
    \label{fig:lombscargle_cubic}
\end{figure}

The Lomb-Scargle periodogram (\cite{1976Ap&SS..39..447L}; \cite{1982ApJ...263..835S}) is a statistical tool used to detect and characterize periodic signal over time in a data set. Compared to other methods such as Fourier transforms, the Lomb-Scargle method can be used on unevenly sampled data, and works by fitting sinusoids over a range of trial frequencies. For each frequency, it measures how well a sinusoidal model fits the data, producing a \say{power} that indicates the strength of the periodic signal. The frequency with the highest power is considered the most likely periodicity present in the data, and gives us the \say{best frequency} \citep{2018ApJS..236...16V}. The best period is then computed easily as $\frac{1}{best\text{ } frequency}$. We use the Lomb-Scargle method to characterize the oscillations of the residuals in CTE-loss around the fit lines, as seen in Figures \ref{fig:lombscargle_linear}, \ref{fig:lombscargle_quad}, and \ref{fig:lombscargle_cubic}. 

To estimate uncertainty on the computed \say{best periods}, we randomly resample the data (time and residuals) and recompute the Lomb-Scargle periodogram each time. We collect the best period from each resample and compute the standard deviation of these periods, which we use as the uncertainty in the best period.

\clearpage

\clearpage
 
\end{document}